\documentclass[aps,prl,showpacs,twocolumn,floatfix,superscriptaddress]{revtex4-1}
\usepackage{epsfig,graphics,color}
\usepackage{graphicx}
\usepackage{dcolumn}
\usepackage{bm}
\usepackage{amsmath,eufrak,mathrsfs}

\newcommand \kd  {\delta}

\newcommand \g {\gamma}

\newcommand \x {\cdot}

\newcommand \A {\alpha}

\newcommand \lc {\langle}
\newcommand \rc {\rangle}

\newcommand \bvec{\left( \begin{array}{c} }
\newcommand \evec{\end{array} \right)}

\newcommand \bea{\begin{eqnarray} }
\newcommand \eea{\end{eqnarray} } 
\newcommand \nn {\nonumber}
\newcommand {\be} {\begin{equation}}
\newcommand {\ee} {\end{equation}}
\newcommand {\epem} {$e^+ e^-$}
\newcommand {\mbx} {\mbox{}}

\newcommand {\ata} {& \times &}

\begin{document}

\title{The $x$ and $Q^2$ dependence of $\hat{q}$, quasi-particles and the JET puzzle}

\author{Evan~Bianchi}
\affiliation{Department of Physics and Astronomy, Wayne State University, Detroit, MI 48201.}

\author{Jacob~Elledge}
\affiliation{Department of Physics and Astronomy, Arizona State University, Tempe, AZ 85287.}

\author{Amit~Kumar}
\affiliation{Department of Physics and Astronomy, Wayne State University, Detroit, MI 48201.}

\author{Abhijit~Majumder} 
\affiliation{Department of Physics and Astronomy, Wayne State University, Detroit, MI 48201.}

\author{Guang-You~Qin}
\affiliation{Institute of Particle Physics and Key Laboratory of Quark and Lepton Physics (MOE), 
Central China Normal University, Wuhan, 430079, China}
\affiliation{Department of Physics and Astronomy, Wayne State University, Detroit, MI 48201.}

\author{Chun~Shen}
\affiliation{Department of Physics, Brookhaven National Laboratory, Upton NY 11973.}

\date{\today}

\begin{abstract} 
We present the first attempt to extract the ``$x$'' dependence of the parton distribution function (PDF) of the quark gluon plasma (QGP). 
In the absence of knowledge regarding the mass of a QGP constituent, we define the new variable $x_N$, 
the nucleonic momentum fraction, 
which represents the ratio of the momentum of the parton to that of a self-contained section of the plasma that has the mass of a nucleon. 
Calculations are compared to data for single hadron suppression in terms of the nuclear modification factor $R_{AA}$ and the azimuthal anisotropy parameter $v_{2}$, as a function of transverse momentum $p_{T}$, centrality and energy of the collision.  
It is demonstrated that the scale evolution of the QGP-PDF is responsible for the reduction in normalization of the jet transport coefficient $\hat{q}$, between fits to Relativistic Heavy-Ion Collider (RHIC) and Large Hadron Collider (LHC) data; a puzzle, first discovered by the JET collaboration.  
Best fits to data are suggestive of the presence of quasi-particles with wide dispersion relations in the QGP.

\end{abstract}

\maketitle

Full jet and intra-jet observables in heavy-ion collisions~\cite{Appelshauser:2011ds,*Cole:2011zz,*Tonjes:2011zz,*Chatrchyan:2012gw,*Aad:2014wha} currently suffer from a series of uncertainties: 
The effect and modeling of hadronization in the presence of the medium, 
the effect and calculation of four-momentum deposition in the medium and reconstructed within the jet etc.
In contrast, leading hadron observables stand on a much firmer theoretical footing. 
In the limit of hard (high $p_{T}$) hadrons, one may apply the methods of factorized pQCD~\cite{Collins:1985ue,*Collins:1988ig,*Collins:1989gx}, 
which factorize the hard partonic modes within the jet from the final fragmentation function [$D(z,Q^{2})$] 
(there is now sufficient experimental evidence that for $p_T > 8$GeV, at RHIC or LHC, the fragmentation is vacuum like), 
the initial Parton Distribution Functions (PDFs), denoted as $G(x,Q^{2})$, and jet transport coefficients in the dense medium, 
such as $\hat{q}$ (the transverse momentum diffusion coefficient)~\cite{Baier:2002tc} and $\hat{e}$ (the longitudinal momentum drag coefficient)~\cite{Majumder:2008zg}. 
These soft matrix elements are universal parametrized functions, extracted from other experiments such as 
\epem~annihilation [$D(z,Q^{2})$] 
and Deep-Inelastic scattering [$G(x,Q^{2})$], 
or set using one or two data points from leading hadron suppression in heavy-ion collisions (e.g. $\hat{q}$). 
Even though there is no factorization theorem established for heavy-ion collisions, 
the calculation of high $p_T$ ($>8$ GeV) leading hadron suppression constitutes one of the more rigorous calculations of 
jet modification in heavy-ion collisions, simply due to the large separation of scales between medium and detected hadron.

Beyond this, all Monte-Carlo simulations of energy loss are, in 
some form or another, dependent on an energy loss formalism~\cite{Wiedemann:2000tf,*Wang:2001ifa,*Gyulassy:2000er,*Arnold:2002ja,*Majumder:2010qh}. 
Hence, straightforward calculations of leading hadron suppression from the underlying energy loss formalism must reproduce 
all available data on single hadrons prior to applications to more sophisticated full jet observables. 
In this paper, 
we report on the results of this test applied to the multiple scattering, multiple emission higher-twist calculation of energy loss~\cite{Majumder:2007hx,*Majumder:2007ne,*Majumder:2009zu,*Majumder:2009ge},
in comparison with a wide variety of available data from RHIC and LHC, on light flavor suppression. 

A simplified version of this test was applied to five different formalisms of energy loss in the recent publication by the JET collaboration~\cite{Burke:2013yra}. 
All the differing energy loss formalisms were applied to calculate the single hadron suppression (expressed by the nuclear modification factor $R_{AA}$, see below) 
for only the most central (0-5\%) events at RHIC and LHC collisions. All were run on an identical hydro-dynamical simulation: the 2+1D viscous simulation of 
Ref.~\cite{Shen:2014vra}. While their remained several differences among the formalisms, in the implementation of the initial hard scattering, energy loss in the hadronic state etc., there 
was across-the-board agreement on the fact that the interaction strength, 
\bea
\hat{\mathscr{Q} }= \frac{\hat{q}(T)}{T^3},
\eea
at the same temperature, is lower at the LHC than at RHIC (we refer to this as the JET puzzle).

This led the authors of Ref.~\cite{Xu:2015bbz} to propose a non-monotonic dependence of both  $\eta/s$ and  $\hat{\mathscr{Q}}$ as functions of temperature: 
Coupled with a dip in $\eta/s$, they also required 
an upward cusp in $\hat{\mathscr{Q}}$ in the region around $T_c$ (based on the quasi-particle relation derived in Ref.~\cite{Majumder:2007zh}). 
Experiments at RHIC are more sensitive to this rise in $\hat{\mathscr{Q}}(T)$ near $T_c$, due to lower initial temperatures 
at RHIC, compared to the LHC. 
As a result, the effective $\hat{\mathscr{Q}}$ extracted in comparison with data tends to be higher at RHIC than at LHC.
 
In this Letter, we challenge the proposal of Ref.~\cite{Xu:2015bbz}, by demonstrating that at both RHIC and LHC, 
both the centrality dependence of the $R_{AA}$ and that of the 
azimuthal anisotropy ($v_2$) of leading hadrons can be described using a $\hat{\mathscr{Q}}$, 
that has no such cusp like behavior near $T_c$ (in this effort, similar to Ref.~\cite{Burke:2013yra}, we will ignore the minor effect of $\hat{e}$~\cite{Qin:2012fua,*Qin:2014mya} on light flavors). 
Note that the initial temperature in a heavy-ion collision (at thermalization) not only depends on the energy of the collision, 
but also on its centrality, with more peripheral collisions having a lower initial temperature distribution. 
The effect of the cusp at $T_{c}$ should be much stronger in such collisions, leading to noticeably larger suppression than expected based on a 
monotonic scaling relation between $\hat{q}$ and $T$. However, no such effect has been found in the current work. 
This is consistent with recent observations within the ASW formalism as reported in Ref.~\cite{Andres:2016iys}.

In this Letter, we demonstrate that the difference in normalization between RHIC and LHC can be straightforwardly understood based on the scale evolution of $\hat{q}$. 
 Based on most fluid dynamical simulations, the highest initial temperature at the LHC is a bare 20\% higher than at RHIC. 
 However, the energy of the jets at the LHC (hadron $p_{T}\sim$100~GeV) is an order of magnitude higher than those of RHIC (hadron $p_{T}\sim$10~GeV). This also 
 implies that leading partons have a considerably (though logarithmically) higher virtuality at the LHC, and as a result, a smaller transverse size.
 At these short distance scales, probed by the jet, the 
 QGP may appear more dilute (unless the energy of the jets is very high). This effect is similar to the downward evolution of PDFs at finite values of $x$, with increasing $Q^{2}$~\cite{Gribov:1972ri,*Gribov:1972rt,*Altarelli:1977zs,*Dokshitzer:1977sg}.  

To estimate the effect of scale evolution on the resolution of the medium, 
we start by considering the analogous process of Deep-Inelastic Scattering on a large nucleus ($A \gg 1$) 
at large photon virtuality ($Q^{2} \gg \Lambda_{QCD}^{2}$) and 
focus on the limit where a hard quark is produced. In this limit, one may factorize the propagation of the quark 
in the extended medium from the production process, 
obtaining equations for the scattering induced single gluon emission spectrum or the transverse momentum distribution of the produced quark. 
In the Breit frame, 
the virtual photon $\g^*$ and the nucleus have 
momentum four vectors $q =\left[ -Q^2/2q^-, q^-, 0, 0\right]$ and $  
P_A \equiv A[p^+,0,0,0]$. 
Where the photon invariant $q^{2}$ is written as $-Q^{2}$.
In this frame, the Bjorken variable is obtained as 
$x_B = Q^2/2p^+q^-$. The momentum of a quark or gluon in any of the 
nucleons is given as $p_{q,g} \simeq xp$ (where $0<x<1$).

The simplest process where $\hat{q}$ appears is the differential hadronic tensor to produce a hard parton with momentum $\l_{\perp}$ transverse to the 
direction of the originating hard jet, caused by single scattering on a nucleon down the path of the jet, 
\bea
\mbx \!\!\!\!\!\frac{dW^{\mu \nu}}{d^{2} l_{\perp}} \!\! &=& \! C^{A}_{p_{A} p_{B}}  
W_{0}^{\mu \nu} \left[ \nabla_{l_{\perp}}^{2}  \kd^{2}(\vec{l}_{\perp})  \right]   
\int\limits_{0}^{L^{-}} d Y^{-}  \left[  \frac{4 \pi^{2} \A_{s}}{2 N_{C}} \right.  \nn \\
\ata  \int \frac{dy^{-}}{2 \pi} \int \frac{d^{2} k_{\perp}  d^{2} y_{\perp}}{(2 \pi)^{2}} 
e^{ -i \frac{k_{\perp}^{2}}{2 q^{-}} y^{-} + i \vec{k}_{\perp} \x  \vec{y}_{\perp} }  \nn \\
\ata \left. \lc p_{B} |  {F^{a}}^{+ \alpha} (Y^{-} \!\!\!+ y^{-}, \vec{y}_{\perp}) {F^{a}}^{+}_{\alpha} (Y^{-}) | p_{B} \rc  
\frac{\mbx^{\mbx}}{\mbx_{\mbx}}\right] \nn  \\
&=& \!  C^{A}_{p_{A} p_{B}}  
W_{0}^{\mu \nu} \left[ \nabla_{l_{\perp}}^{2}  \kd^{2}(\vec{l}_{\perp}) \right]   
\int\limits_{0}^{L^{-}} d Y^{-}  \hat{q}(Y^{-}). \label{single-scattering}
\eea
In the equation above, $W^{\mu \nu}_{0}$ represents the case where there is no 
rescattering of the produced quark. The factor $C^{A}_{p_{A} p_{B}}$ represents the probability to find a nucleon with 
a momentum $p_{A}$ in the nucleus with $A$ nucleons and its possible correlation with a nucleon with momentum $p_{B}$ 
further within the nucleus. Without any rescattering, the $\vec{l}_{\perp}$ distribution is given approximately by a $\kd$-function.
With one rescattering, the $\kd$-function is shifted by $\int d Y^{-} \hat{q}$. 
Further uncorrelated multiple scattering results in the diffusion equation described in 
Ref.~\cite{Majumder:2007hx}. 

\begin{figure}[htbp]
\resizebox{3.1in}{1.75in}{\includegraphics{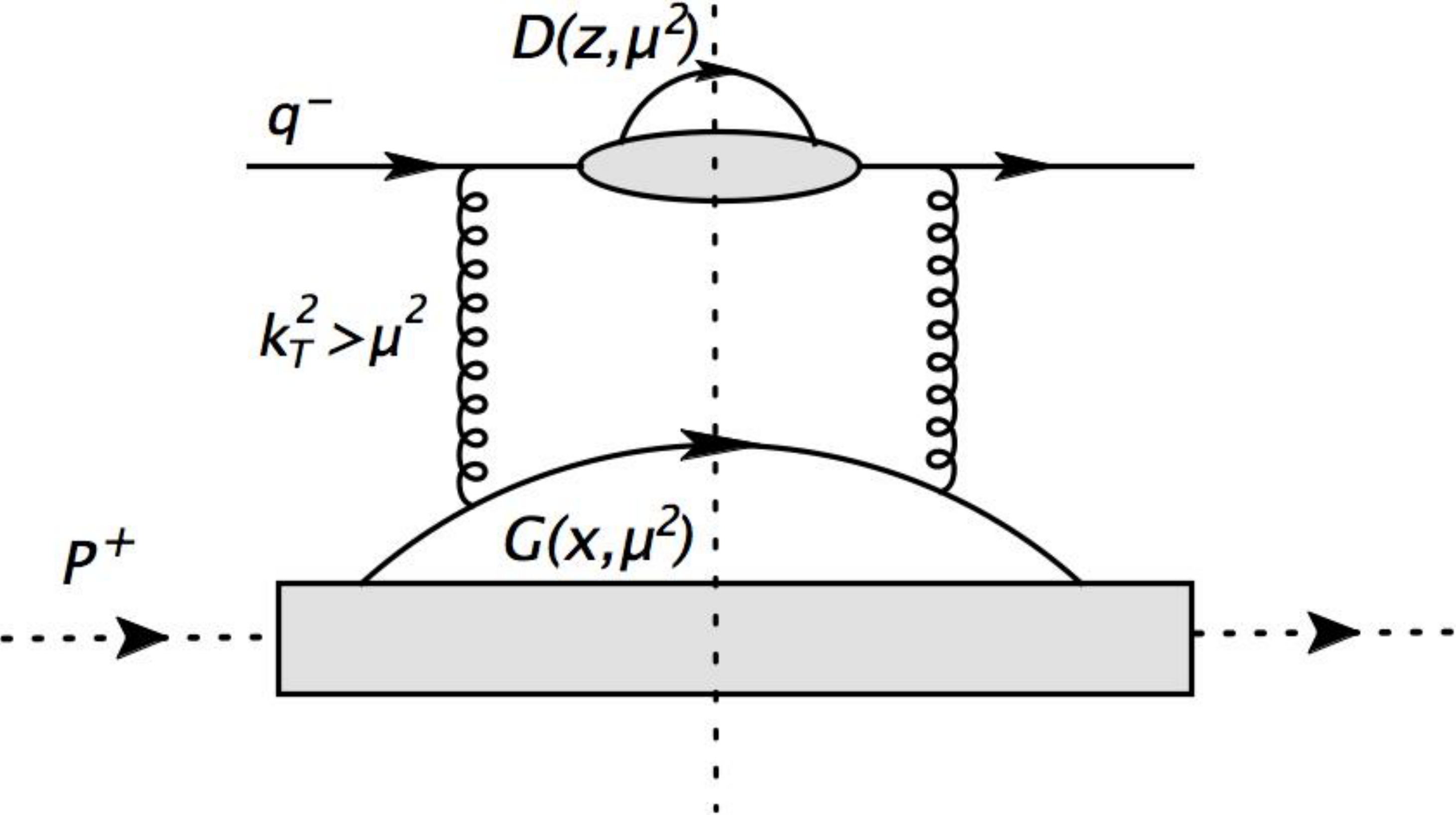}} 
    \caption{A hard quark with large light-cone momentum $q^-$ scattering via a Glauber gluon $k_\perp $with a hard collinear fluctuation in a target constituent (nucleon in nucleus or QGP constituent). }
    \label{qhat}
\end{figure}

In order to obtain the energy and scale dependence, we consider the correlator that defines $\hat{q}$: Eq.~\eqref{single-scattering} shows that, 
unlike a collinear factorized parton distribution function, the leading light-cone 
momentum $k^{+}$ is not independent of the transverse momentum $k_{\perp} $. Indeed, they are related as 
$k^{+} = k_{\perp}^{2} / 2q^{-}$. 
In this sense, it is more similar to the case of the 
$k_{T}$ factorized unintegrated gluon distribution in high energy [or Balitsky-Fadin-Kuraev-Lipatov (BFKL)] 
evolution~\cite{Fadin:1975cb,*Kuraev:1977fs,*Balitsky:1978ic}. However, estimating the $x = k_{\perp}^2/ (2 p^+ q^-) \simeq \hat{q} L/ (2 M E)$, where $M$ is the  
mass of the proton, and $E$ is the energy in the rest frame of the nucleus, 
one obtains $x \geq 1 \times 10^{-3}$ in a large nucleus, for the lowest extracted value of $\hat{q} L \sim 0.1$GeV$^2$~\cite{Chang:2014fba}, and a 
parton energy of $E \sim 50$ GeV. This range of $x$ is argued to be outside the regime of the BFKL equation~\cite{Abramowicz:2015mha}. 

\begin{figure}[htbp]
\resizebox{3.1in}{3.3in}{\includegraphics{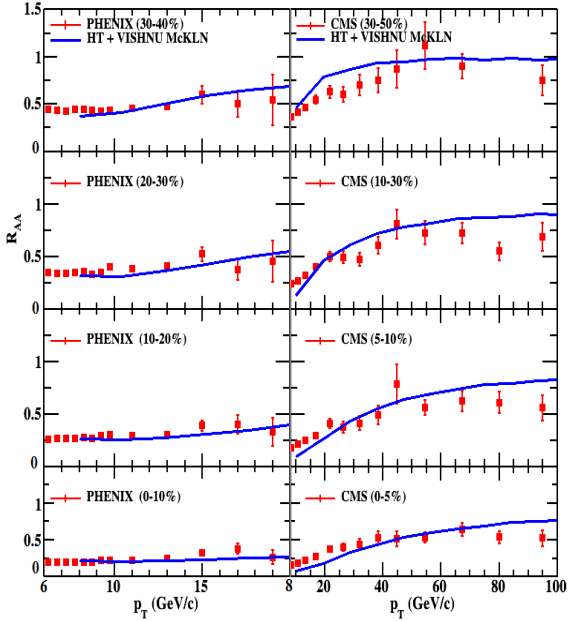}} 
    \caption{ Nuclear Modification factor at two different collision energies for four different centralities at RHIC and LHC. The parameters of the QGP-PDF are dialed to fit the bottom two panels.}
    \label{Raa}
\end{figure}

Extending this argument to jets traversing through a QGP is not without caveat: we define a ``nucleonic'' $x_N = \hat{q} L/ (2 M_N E) $, where $M_N$ is the mass of a nucleon ($=1$GeV in the following), 
and $E$ is the energy of the jet in the rest frame of the struck portion of QGP. 
We define this quantity due to a lack of knowledge of $m_{QGP}$, the mass of a degree of freedom, or a correlated enclosure within a QGP; 
all estimates of this mass place it at $m_{QGP} \sim gT \lesssim M_{N}$. Thus at best, our $x_N = \xi (T) x$, 
where $x$ is the actual momentum fraction of a parton within a QGP degree of freedom, and $0<\xi(T)<1$ is a temperature dependent scaling factor. 
For cases where $x_N$ is evaluated in a QGP, the $\hat{q}$ can be anywhere between 50 to a 
100 times higher than in nucleus and thus $x_N \gtrsim 5 \times 10^{-2}$.  
It is possible that at higher energies, BFKL like effects will become important~\cite{Abir:2015qva,*Iancu:2014kga,*Blaizot:2014bha}. However, as we will demonstrate, for 
most jet energies at RHIC and LHC, a DGLAP based formalism, as described below will provide a better description of the experimental data.

For intermediate values of $x_N$, the evolution of $\hat{q}$ is obtained from the diagram in Fig.~\ref{qhat}: Partonic 
fluctuations (indicated by solid lines with arrows), which may be a quark or a gluon at a scale $\mu^2$, collinear with the target state of a nucleon or a QGP constituent (with target momentum $P^+$), radiate 
a Glauber gluon~\cite{Idilbi:2008vm}, with resolved transverse momentum $k_\perp^2 \gtrsim \mu^2$, which scatters off the outgoing jet parton, with momentum $q^-$, which in turn 
fragments at the scale $\mu^2$. The upper limit of the exchanged transverse momentum $k_\perp^2 < 2 x_N P^+ q^- $ is given by the kinematic bound.  
This 
exchange introduces scale and energy dependence within 
the transport coefficient $\hat{q}$. A similar concept for the evolution of $\hat{q}$ has also been proposed in Refs.~\cite{Kang:2014ela,*CasalderreySolana:2007sw}.

\begin{figure}[htbp]
\resizebox{3.1in}{3.3in}{\includegraphics{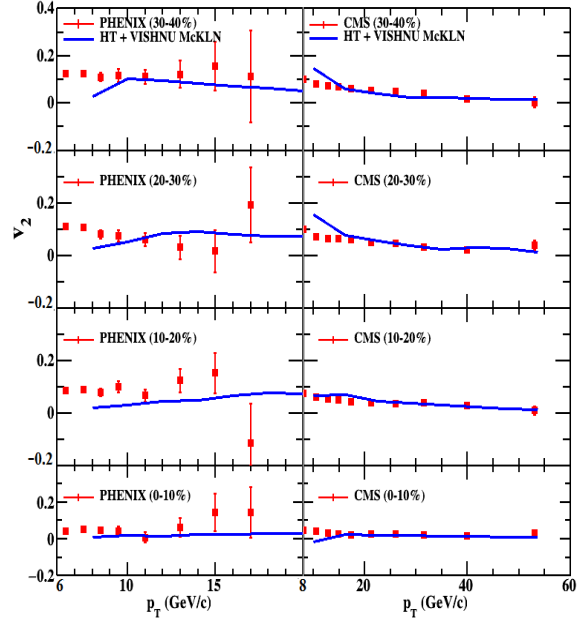}} 
    \caption{Azimuthal anisotropy at two different collision energies for four different centralities at RHIC and LHC.}
    \label{v2}
\end{figure}

The input to such a calculation is the form of the PDF ($x_N$ dependence) within the QGP at the low scale of 1~GeV. The PDF at any higher scale is 
obtained by DGLAP evolution. For this first attempt, we parametrize the PDF in standard form:
\bea
G(x,\mu^2=1GeV^2) = N x^\alpha (1-x)^\beta.
\eea
Where, $N, \alpha$ and $\beta$ are set by best fits to the data. The three coefficients are not independent of each other: The choice of $\alpha$ and 
$\beta$ restricts the choice of $N\!,$ which now replaces the overall normalization $\hat{q}_0$. 
Thus, there are effectively 2 input parameters that have to be fit to data; in spite of the introduction of scale and energy dependence, 
there is no increase in the number of input parameters, as 
compared to current calculations without evolution in $\hat{q}$, which normalize $\hat{q}_0$ at RHIC and LHC energies separately. 

In this Letter, we present results for the event averaged nuclear modification factor 
\bea
R_{AA} = \frac{ \frac{ d^{2} N_{AA} ( b_{min}, b_{max} )  }{d^{2}p_{T} dy }   } 
{  \langle N_{bin} ( b_{min}, b_{max} ) \rangle  \frac{d^{2} N_{pp} }{ d^{2 }p_{T} dy } },
\eea
where  $d^{2} N_{AA}(b_{min}, b_{max})$ is the differential yield of hadrons in 
bins of $p_{T}$, rapidity ($y$) and centrality (codified by a representative range of impact parameters 
$ b_{min}$ to $b_{max}$). This is divided by the differential yield in a $p$-$p$ collision, scaled by 
$\langle N_{bin} \rangle$, the mean number of binary nucleon-nucleon collisions in the
same centrality bin. Our calculations closely follow the methodology of 
Ref.~\cite{Majumder:2011uk}, which calculates the $R_{AA}$ given a $\hat{q}_0$, assuming that $\hat{q}$ scales with the local entropy density ($\hat{q} = \hat{q}_{0} s / s_{0}$, where $s_{0}=96$/fm$^{3}$). The main difference 
in the current work is the scale and energy dependence of $\hat{q}_0$. 
This evolution causes a reduced $\hat{q}$ at most values of 
$x_N$ probed, leading to a natural reduction in the mean value of $\hat{q}/T^3$ at LHC compared to RHIC.

\begin{figure}[htbp]
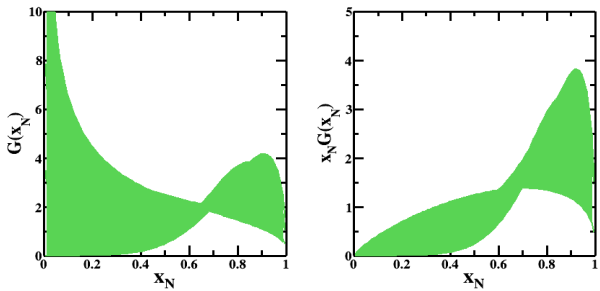

\resizebox{1.5in}{1.5in}{\includegraphics{fx.pdf}} \hspace{0.2cm}\resizebox{1.5in}{1.5in}{\includegraphics{xfx.pdf}} 
    \caption{Left: The input PDF [$G(x_N)$] at $\mu^2=1$~GeV$^2$, used in the calculation of the scale dependence of $\hat{q}$. Right: $x_N$ times the PDF. The band in both plots represents 
    the range of functions that yield very similar $\chi^{2}$/d.o.f. when compared with Fig.~\ref{Raa}.}
    \label{InputPDF}
\end{figure}

We vary $\alpha$ and $\beta$ to obtain the combined best fit for the 0-5\% centrality bin at the LHC 
and the 0-10\% centrality bin at RHIC, as these have the 
smallest error bars. The variation with $p_T$, centrality, and now also $\sqrt{s}$ of the collision are predictions. 
The results from this particular choice of $\alpha$ and $\beta$ for 4 different centralities is presented in Fig.~\ref{Raa} for both RHIC and LHC energies.
We stress once again that there is no re-normalization between RHIC and LHC energies. The reduction in the interaction strength $\hat{\mathscr{Q}} = \hat{q}/T^3$ is 
entirely caused by scale evolution in $\hat{q}$. We also compare the azimuthal anisotropy $v_2$ from the angle dependent $R_{AA}$, for 4 different centralities at RHIC and LHC energies in Fig.~\ref{v2}, where 
\bea
\mbox{}\!\!\!\!R_{AA} (p_T,\phi) = R_{AA} \left[  1 + 2 v_2 \cos( 2 \phi - 2 \Psi)  + \ldots \right].
\eea
where $\Psi$ is the event plane angle determined by the elliptic flow of soft hadrons. 
As the quenching of jets is carried out on a ``single-shot'' or event averaged hydro calculation, there is a well defined event plane in these fluid dynamical simulations, 
One should note that Fig.~\ref{v2} represents a parameter free calculation. All input parameters have been set in the angle integrated $R_{AA}$ calculations presented in Fig.~\ref{Raa}.

The improved fit with experimental data ($\chi^2$/d.o.f=5.6 for Fig.~\ref{Raa}) without the need for an arbitrary renormalization of $\hat{\mathscr{Q}}$ between RHIC and LHC energies indicates that scale dependence of $\hat{q}$ represents an actual physical effect for jets traversing a QGP. 
This fit also adds confidence in the input PDF at $\mu_0^2 = 1$~GeV$^2$, within a QGP constituent. We have attempted several values of $\alpha$ and $\beta$ and 
obtained a shallow minimum in $\chi^2$/d.o.f. This allows us to isolate the input distribution to lie within the bands in Fig.~\ref{InputPDF}. The isolated range of input 
distributions has a ``valence'' like bump around $x_N\!\sim\!0.8$ and a large ``sea'' like contribution at small-$x_N$. The wide bump around $x_N\!\sim\!0.8$ would be consistent 
with that obtained from a plasma of quasi-particles.

In this Letter, we have presented the first successful attempt to explain the JET puzzle: the downward renormalization of the interaction strength $\hat{\mathscr{Q}}$, at 
the same temperature, at LHC energies compared to RHIC energies. This was achieved by allowing $\hat{q}$ to vary with the scale of the jet. This scale dependent $\hat{q}$ 
was obtained by considering the scattering of a hard parton off Glauber gluons radiated off a target PDF evolved from an input distribution up to the scale of the jet parton.
Fits with experimental data on $R_{AA}$ and $v_2$ at 4 different centralities from both RHIC and LHC were presented.
The inferred input distribution at the lowest possible perturbative scale of 1~GeV$^2$, has a valence like bump consistent with the presence of quasi-particles in the QGP. 
The bands in Fig.~\ref{InputPDF} represent the uncertainty in the input PDF. The reduction of uncertainty will require extensive comparisons with experimental data beyond leading 
hadron suppression. These and other details of the calculation may be found in Ref.~\cite{kumar2017}.

\acknowledgements
This work was supported in part by the National Science Foundation (NSF)
under grant number PHY-1207918, by the U.S. Department of Energy (DOE)  
under grant number DE-SC0013460 and the NSFC under grant number 11375072. 
The work of E.B. and J.E. was supported by the Wayne State REU (NSF grant PHY-140853).

\bibliography{refs}

\end{document}